# Energy-saving technologies and energy efficiency in the post-pandemic world

*Wadim Strielkowski [1,2]*Larisa Gorina[3], Elena Korneeva[3,4], Olga Kovaleva[5]*

[1] University of California, Berkeley, United States
[2] Czech University of Life Sciences, Prague, Czech Republic
[3] Togliatti State University, Togliatti, Russia
[4] Financial University under the Government of the Russian Federation, Moscow, Russia
[5] Orenburg State University, Orenburg, Russia
Correspondence: E-mail: strielkowski@berkeley.edu

**ABSTRACT**

This paper explores the role of energy-saving technologies and energy efficiency in the post-COVID era. The pandemic meant major rethinking of the entrenched patterns in energy saving and efficiency. It also provided opportunities for reevaluating energy consumption for households and industries. In addition, it highlighted the importance of employing digital tools and technologies in energy networks and smart grids (e.g. Internet of Energy (IoE), peer-to-peer (P2P) prosumer networks, or the AI-powered autonomous power systems (APS)). In addition, the pandemic added novel legal aspects to the energy efficiency and energy saving and enhanced inter-national collaborations and partnerships.

The paper highlights the importance of energy efficiency measures and examines various technologies that can contribute to a sustainable and resilient energy future. Using the bibliometric network analysis of 12960 publications indexed in Web of Science databases, it demonstrates the potential benefits and challenges associated with implementing energy-saving technologies and autonomic power systems in a post-COVID world. Our findings emphasize the need for robust policies, technological advancements, and public engagement to foster energy efficiency and mitigate the environmental impacts of energy consumption.

## 1. INTRODUCTION

The recent COVID-19 pandemic had far-reaching consequences on global energy consumption patterns, necessitating a re-evaluation of energy use and practices across industries and among consumers. As the world looks towards recovery and resilience in the post-COVID era, the role of energy-saving technologies and energy efficiency becomes increasingly critical (Amir and Khan, 2022; Echegaray, 2021).

The pandemic not only disrupted the global demand for energy but also shed light on the relevance of such innovative trends in energy market as the existence of prosumers and peer-to-peer (P2P) energy networks (Siuta-Tokarska et al., 2022; Kalina, 2023). These emerging trends carry the capability to transform the energy sector, granting consumers the authority to actively engage in both energy generation and distribution processes. Moreover, by embracing energy efficiency measures, such as energy-saving technologies or the novel concept of the autonomic power systems (APS) (self-configuring, self-healing, self-optimizing and self-protecting, decentralized, low-level intelligence autonomous systems making the decisions necessary to meet the priorities of the system's stakeholders), societies can strive for a more sustainable future (Strielkowski, 2017; Strielkowski et al., 2019a; Das et al., 2021).

The COVID-19 pandemic brought about unprecedented challenges and disruptions across the globe. As countries gradually recover and rebuild their economies, there is a growing realization that energy efficiency must be at the forefront of our efforts to create a sustainable and resilient future (Karakosta et al., 2021; Kaklauskas et al., 2021). The post-COVID era presents us with an opportunity to rethink our approach towards energy consumption and embrace innovative energy-saving technologies (Lee & Woo, 2020). The global health crisis has brought to the forefront the weaknesses in our existing systems, underscoring the necessity for optimizing resource utilization and efficiency. Energy efficiency not only reduces greenhouse gas emissions but also helps in reducing costs, enhancing energy security, and mitigating climate change (Nielsen et al., 2021). One of the key lessons learned from the pandemic is that working remotely can significantly reduce energy consumption associated with commuting and operating commercial buildings (Mantesi et al., 2022; Navaratnam et al., 2022). As businesses adapt to new work models, it becomes crucial to explore ways to optimize energy usage in both residential and commercial settings (Godina et al., 2020). Moreover, as governments implement economic recovery plans, investing in clean and efficient technologies can create job opportunities while simultaneously reducing environmental impacts (Tian et al., 2022). By prioritizing energy efficiency measures in sectors such as transportation, manufacturing, and construction, we can foster sustainable growth while minimizing the carbon footprint of society and economy (Singh et al., 2023).

Our review conducts a bibliometric analysis of the publications focusing on energy efficiency and energy saving in the pre-COVID and post-COVID world using both the Google Trends toolkit and the VOSViewer software for scrutinizing text and bibliometric data. By doing this, it attempts to assess the significance of energy-related technologies in the world facing many recent challenges – from technological revolution to the health crises. The uniqueness and additional value of this study lie in its focus on the significance of energy efficiency measures and its exploration of various technologies that can lead to energy conservation and foster sustainability in the post-COVID era. Through an analysis of case studies, papers, and policy reports, we explore the potential benefits and challenges associated with implementing energy-saving technologies and autonomic power systems. By understanding these nuances, one can identify strategies and best practices to overcome barriers and maximize the potential of these technologies. Our work is aimed at providing insights into the role of energy-saving technologies and energy efficiency in the post-COVID era. It aims to contribute to the growing body of knowledge sur-rounding sustainable energy practices.

## 2. LITERATURE REVIEW

### 2.1. Impacts on the transportation, industry, and residents

To put it mildly, the COVID-19 pandemic cast a significant blow to energy consumption worldwide. As countries implemented lockdown measures, industries and businesses were forced to shut down or operate at reduced capacity, leading to a decrease in energy demand (Soava et al., 2021; Gorina et al., 2023). As people had to isolate themselves to slow down the spread of the virus, residential energy usage changed its patterns. For example, energy consumption related to transportation declined. Restrictions on travel and remote working policies resulted in fewer cars on the road and reduced air travel. Consequently, the demand for gasoline and jet fuel plummeted, leading to lower emissions from these sectors (Jiang et al., 2021; Munawar et al., 2021). Additionally, as commercial buildings remained vacant during lockdowns, electricity consumption for offices, retail spaces, and other non-residential structures declined significantly due to the reduction in HVAC system operational needs (Kikstra et al., 2021).

At the same time, energy usage in the residential sector increased due to the fact that people were spending more time indoors and the need for heating and cooling spiked. Moreover, with individuals relying heavily on electronic devices for work from home arrangements and entertainment purposes during lockdowns, household electricity demand surged. Thence, the pandemic offered valuable insights into how changes in human behavior can influence energy consumption patterns (Lőrincz et al., 2021; Proedrou, 2021).

### 2.2. Challenges and opportunities of RES

It is quite obvious that in the past decade or so RES has become synonymous for energy efficiency. They are playing an increasingly leading role in achieving sustainability and resilience (Seferlis et al., 2021; Simionescu et al., 2022). As nations strive to recover from the pandemic's economic and environmental impacts, renewable energy emerges as a crucial component of their strategies. A notable benefit of renewable energy sources lies in their inherent capacity to reduce greenhouse gas emissions (Balali & Stegen, 2021). Unlike fossil fuels, renewables like solar, wind, hydro, and geothermal power produce electricity without the need for burning fossil fuels or releasing detrimental pollutants into the atmosphere. Moreover, integrating renewable sources into existing energy grids enhances overall system efficiency. These technologies produce electricity closer to the point of consumption, minimizing transmission losses associated with long-distance power transport (Jafarinejad et al., 2023; Strielkowski et al., 2019).

Decentralized generation also improves grid resilience by reducing vulnerability to disruptions caused by natural disasters or cyber-attacks. Renewables also provide opportunities for local job creation and economic growth. Allocating resources to renewable technologies fosters innovation, spurs market competition, and simultaneously generates employment prospects across diverse sectors, including manufacturing, installation, maintenance, and research. Embracing renewable energy sources plays a pivotal role in realizing energy efficiency objectives in the post-COVID era (Lu et al., 2022; Panarello & Gatto, 2023; Kaftan et al., 2023). In the aftermath of the COVID-19 pandemic, there has been an increasing focus on energy-conserving technologies and methodologies. As we enter the post-COVID era, smart home solutions are emerging as a promising avenue to reduce energy consumption and promote sustainability (Shokouhyar et al., 2021). Smart meters and smart appliances are becoming more efficient in monitoring energy usage and informing the consumers about their energy consumption patterns through the smartphone apps (Jabbar et al., 2019; Aliero et al., 2021; Yar et al., 2021). From refrigerators that optimize cooling cycles based on usage patterns to washing machines that adjust water levels according to load size – these innovations contribute significantly towards reducing overall energy consumption. Artificial intelligence (AI) also plays an important part in optimizing energy efficiency for smart homes (Heim et al., 2023; Farzaneh et al., 2021; Al-Khassawneh, 2023).

### 2.3. Prospects and experience

The period following the COVID-19 pandemic offers a distinctive chance to expedite the uptake of energy-conserving technologies and elevate energy efficiency (Luo et al., 2023; Lee & Woo, 2020; De Las Heras et al., 2020). However, several barriers must be overcome to ensure widespread implementation. Firstly, upfront costs often deter individuals and businesses from investing in energy-saving technologies. To address this barrier, governments and financial institutions can provide incentives such as tax breaks, grants, or low-interest loans to make these technologies more affordable (Zhu et al., 2023). (Additionally, raising awareness about the long-term cost savings associated with energy-efficient solutions can encourage individuals and organizations to consider these investments (Liu et al., 2023a). Second, another issue for implementing energy-saving technologies is the lack of technical expertise since many users are now fully aware of their potential and integration. Stakeholders and policymakers need to raise awareness about energy saving options and to popularize these concepts among the general public (Anwar et al., 2020; Mahdavian et al., 2021). Moreover, outdated regulations and policies often hinder the adoption of innovative energy-saving technologies. Governments need to review existing frameworks to eliminate any barriers that prevent the integration of these technologies into buildings or infrastructure projects (Strielkowski et al., 2020; Liao et al., 2023). Streamlining approval processes for implementing such solutions can also facilitate their widespread adoption. Lastly, fostering collaboration between different stakeholders is crucial for overcoming barriers to widespread adoption (Nazari and Musilek, 2023).

Table 1 that follows summarizes the prospects and current experience of the current energy-saving technologies and energy efficiency (see Table 1).

**Table 1.** Current experience and prospects of energy-saving technologies and energy efficiency.

| Sources | Prospects | Current experience |
|---|---|---|
| Stielkowski (2019), Zafar & Ben Slama (2022), Kataray et al. (2023) | Adoption of smart grid systems | Successful implementation in certain regions, enabling better energy management and integration of renewable sources |
| Abdalla et al. (2021), Kabeyi & Olanrewaju (2022), Goudarzi et al. (2022) | Implementation | Growing adoption in commercial and industrial sectors, leading to optimized energy usage and cost savings |
| Farzaneh et al. (2021), Kikstra et al. (2021), Adly and El-Khouly (2022), Liu et al. (2023b) | Utilization of building automation systems | Increasing implementation in smart buildings, improving energy efficiency through automated control of lighting, HVAC, and other systems |
| Jabbar et al. (2019), Yar et al. (2021), García-Quevedo and Jové-Llopis (2021), Muryani et al. (2023) | Optimization of industrial processes | Ongoing efforts to improve energy efficiency in manufacturing and industrial sectors through process optimization and automation |
| Seferlis et al. (2021), Balali & Stegen (2021), Lu et al. (2022), Kaftan et al. (2023), Panarello & Gatto (2023) | Renewable energy sources deployment | Promoting the widespread adoption of renewable energy technologies, such as solar and wind power, to decrease reliance on fossil fuels |
| Woo et al. (2021), Fekete et al. (2012), Yu et al. (2022) | Policy and regulatory measures | Implementation of energy efficiency standards, building codes, and incentives to encourage energy-saving practices |
| Parrish et al. (2020), Jiang et al. (2021), Lőrincz et al. (2021), Čábelková et al. (2022) | Demand response and behavior change | Adoption of demand response programs and initiatives promoting energy-conscious behavior among individuals and organizations |
| Rotta et al. (2019), Khan et al. (2020), Adisorn et al. (2021), Siuta-Tokarska et al. (2022), Gorina et al. (2023), Nazari & Musilek (2023) | Digitalization and data analytics | Integration of digital technologies and data analytics for real-time monitoring and optimization of energy consumption |
| Bertoldi et al. (2012), Rana et al. (2022), Wang et al. (2022), Zhang et al. (2022), Zhao et al. (2022) | Energy efficiency financing and investment | Growing interest in financing energy efficiency projects and investments, driven by financial institutions and government initiatives |
| Rotta et al. (2019), Anwar et al. (2020), Woo et al. (2021), Mahdavian et al. (2021) | Collaboration and knowledge sharing | Increasing collaboration among stakeholders, knowledge exchange platforms, and international partnerships to share best practices |
| Ali et al. (2021), Apollo and Mbah (2021), DellaValle & Czako (2022), Chen et al. (2023) | Research and development efforts | Continued research to identify emerging trends, technologies, and solutions to enhance energy-saving and efficiency |
| Apollo and Mbah (2021), Soava et al. (2021), Gorina et al. (2023) | Education and awareness | Education campaigns and awareness programs to empower individuals and organizations to make informed choices for energy efficiency |
| Hanke et al. (2021), Xie et al. (2022), Dion et al. (2023), Liu et al. (2023) | International cooperation and partnerships | Collaborative initiatives to address global energy challenges, promote sustainable energy practices, and share resources |

### 2.4. Legal aspects of the energy-saving technologies and energy efficiency

Before the onset of the global pandemic, various legal frameworks existed to promote and regulate energy-saving technologies and energy efficiency. These regulations aimed to tackle climate change, reduce greenhouse gas emissions, and enhance sustainability in the energy sector (García-Quevedo & Jové-Llopis, 2021; Ahmed et al., 2020). One key aspect of the pre-pandemic legal framework was the establishment of energy efficiency standards for appliances and equipment (Woo et al., 2021). Many governments-imposed restrictions for businesses to comply with strict energy efficiency standards which include a wide range of measures and financial and tax incentives to adapt them (Economidou et al., 2020; Zhao et al., 2022). These incentives encouraged individuals and businesses to invest in energy-efficient solutions by reducing their upfront costs or providing ongoing financial benefits. Furthermore, building codes and regulations played a crucial role in promoting energy efficiency in construction projects. Governments enacted laws that mandated certain energy-saving measures during building design and construction phases (Zhao & Zhang, 2022). These measures included efficient insulation systems, advanced heating and cooling systems, efficient lighting solutions, and renewable energy integration into buildings. The pre-pandemic legal framework recognized the importance of promoting sustainable practices within industries by implementing emission reduction targets for power plants and industrial facilities. Governments enforced strict emissions standards that required these entities to adopt cleaner technologies or face penalties (Gollakota & Shu, 2023).

The COVID-19 pandemic had a substantial influence on energy consumption trends, underscoring the significance of energy-saving technologies and energy efficiency. As countries strive to recover from the economic downturn caused by the crisis, there has been a renewed focus on sustainable development, leading to changes in legislation and policies related to energy-saving technologies (Proskuryakova et al., 2021; Çelik et al., 2022). One notable change is the introduction of stricter building codes and regulations that prioritize energy efficiency. Governments are now mandating higher standards for insulation, lighting systems, and HVAC in both residential and commercial buildings (Adly & El-Khouly 2022; Liu et al., 2023). The main objective is to promote sustainable energy sources such as solar panels or geothermal systems. In addition, various financial mechanisms are offered that might increase the deployment and acceptance of renewable energy sources, smart grids, and electric vehicles (Bertoldi et al., 2021; Kerstens & Greco, 2023). Embracing clean technology not only contributes to the reduction of greenhouse gas emissions but also fosters the creation of fresh employment prospects within the renewable energy sector (Gielen et al., 2019). Furthermore, post-pandemic legislation is emphasizing the need for greater transparency in reporting energy consumption data. Governments are implementing mandatory disclosure requirements for large energy consumers such as industries or public institutions (Al-Madani et al., 2022). This enables better monitoring of energy usage trends and encourages organizations to implement strategies for reducing their carbon footprint.

## 3. METHODS

In this section, we showcase the outcomes of the bibliometric network analysis focused on research publications encompassing energy-saving technologies and energy efficiency. The analysis was performed using the VOSViewer v. 1.6.15 software, a widely employed tool for identifying prevailing trends in interdisciplinary research across various subjects. Additionally, it helps determine the most pertinent tools and instruments in specific topics, such as energy-saving technologies and energy efficiency, for contemporary academic and scholarly investigations (Strielkowski et al., 2023; Gardanova et al., 2023).

We have chosen the Web of Science Core Collection (WoS) database for our research. Using the keywords "energy efficiency" and "energy saving" in scientific papers, reports, proceedings, and book chapters published between 2018 and 2023 and indexed in WoS database, a sample of 19964 indexed publications was retrieved. Figure 1 that follows provides the summary of data and the algorithm for the data selection in a form of a diagram.

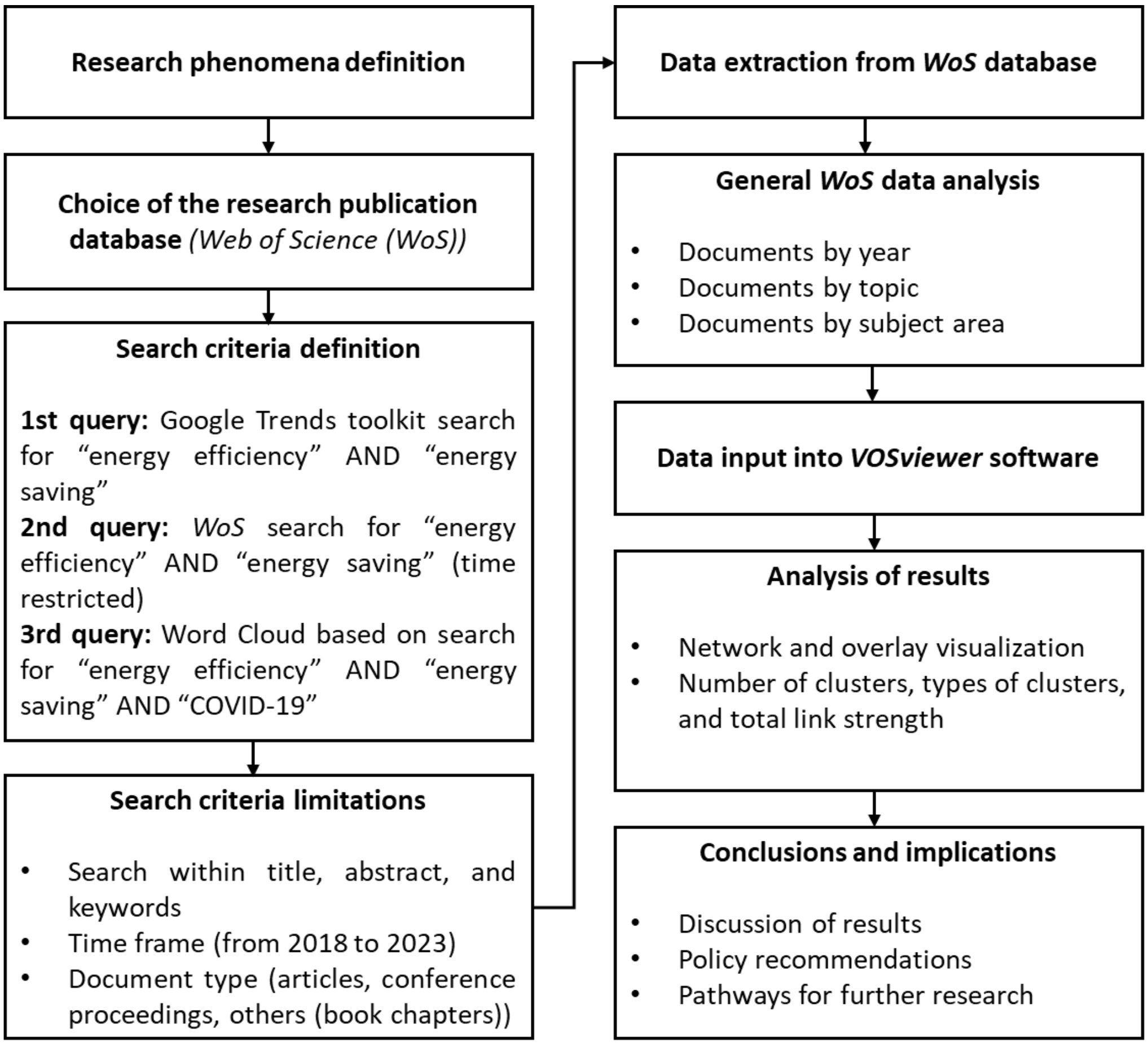

**Figure 1.** Steps of bibliometric data collection and network analysis.

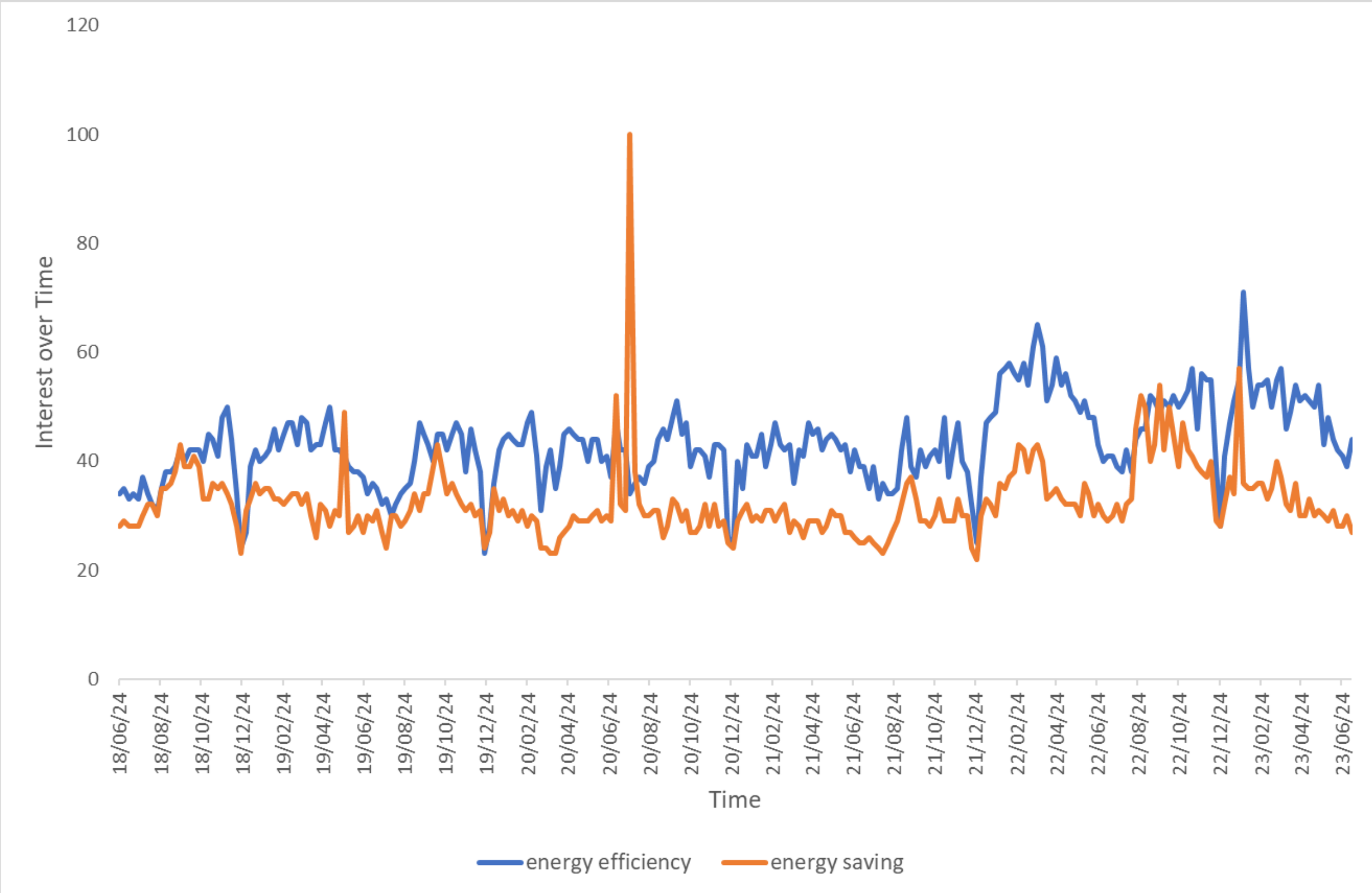

**Figure 2.** Dynamics of frequency of worldwide search requests for keywords "energy efficiency" and "energy saving" (2018–2023).

Before we plunge into the bibliometric analysis, let us first present the results of the keyword searches for "energy efficiency" and "energy saving" in the last five years. Figure 2 that is depicted above indicates the frequency of Internet searches ("interest over time"—the search interest relative to the highest point on the chart for the given region and time) of the keywords "energy efficiency" and "energy saving" worldwide during the past five years (from 2018 to 2023). The figure was created based on the analysis obtained from the Google Trends offered by Google search engine (Google Trends, 2023). Google Trends identifies the change in search queries of the main relevant concepts helping the researchers to assess the peak periods with the periods of the most significant changes in Internet searchers. From the results of this analysis, it becomes quite clear that while the term "energy efficiency" had been a subject of higher search interest in 2018–2023, it was overtaken by the keyword "energy saving" during the first "COVID summer" of 2020 which can probably be explained by the unclear perspectives associated with the worldwide lockdowns, working from home, and facing social isolation.

## 4. RESULTS

Figure 3 that is shown below shows the results of the network map based on the text data from the sample of 12960 publications indexed in WoS database. Our results of the bibliometric network analysis identified four main clusters. Our analysis of the keywords and topics found in the sample of publications retrieved from WoS database revealed that the most frequent concepts were the following: (1) building energy efficiency (see Cluster 1); (2) algorithm and optimization (see Cluster 2); (3) temperature and HVAC (see Cluster 3); as well as (4) production and conversion (see Cluster 4).

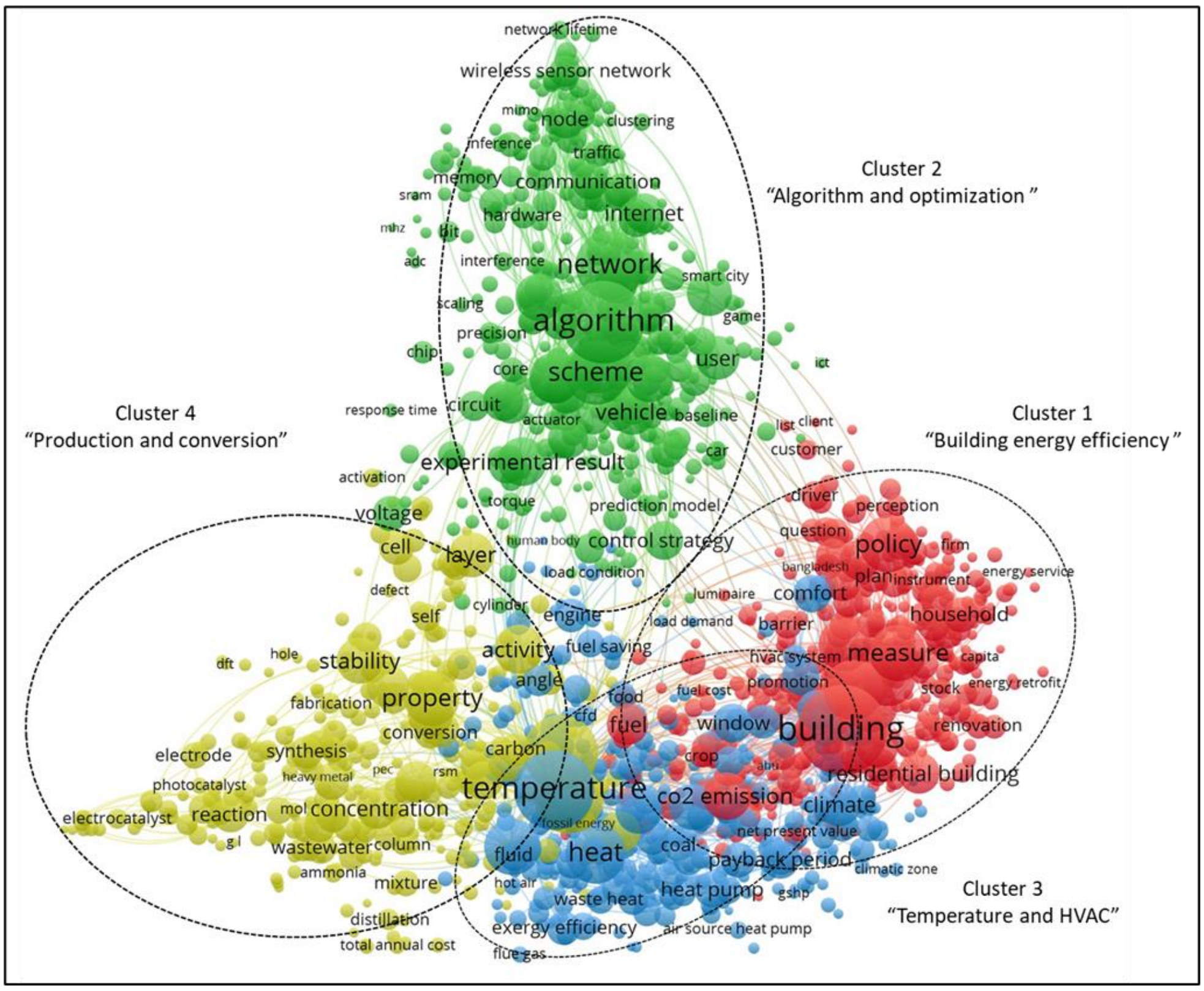

**Figure 3.** The dominant clusters of cross-sector research connected with energy efficiency and energy saving retrieved from the sample of 12960 WoS publications.

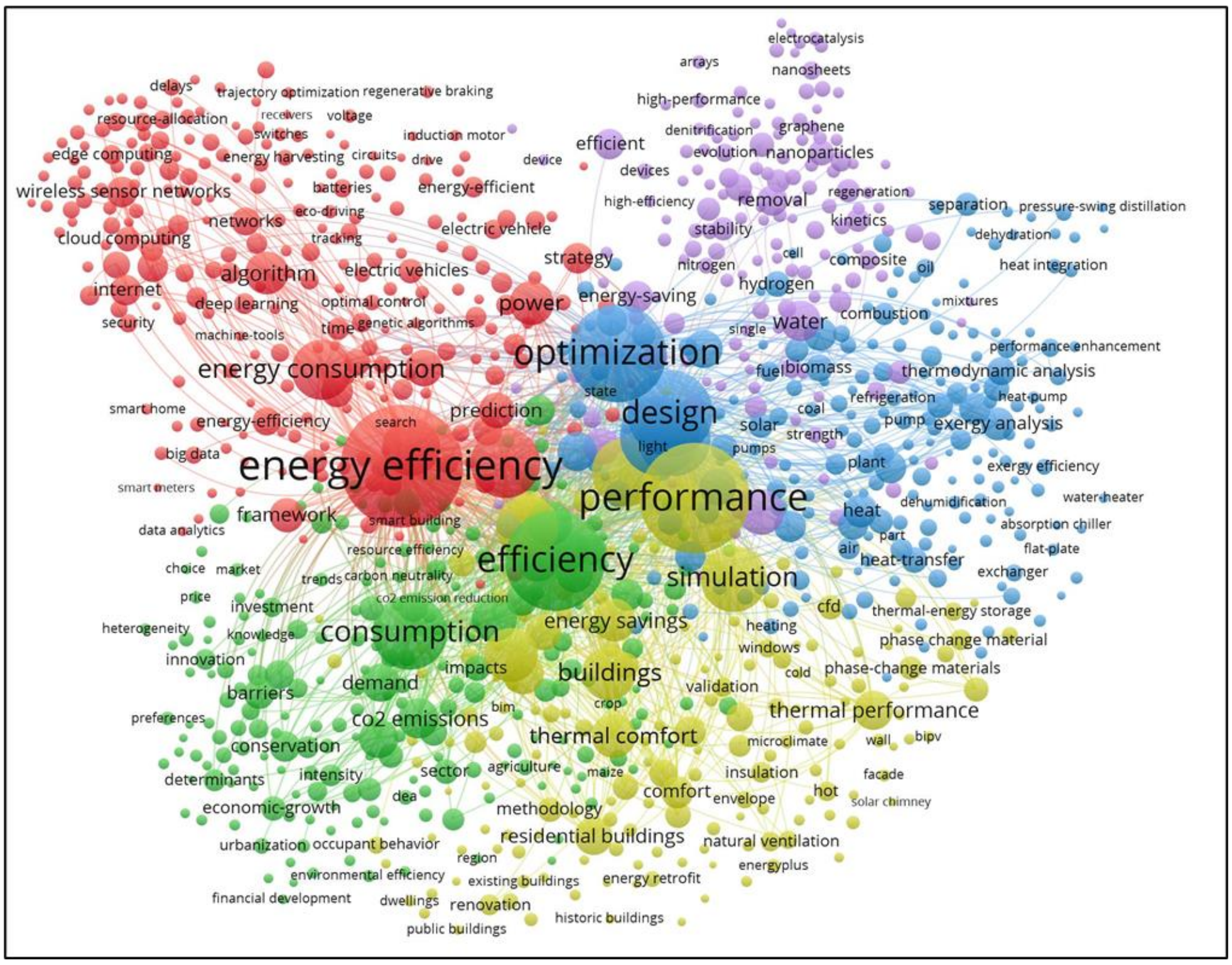

**Figure 4.** Network map based on the bibliographic data of the sample of papers containing the keywords "energy efficiency" and "energy saving" in 12960 publications from WoS database.

Figure 4 above reveals the results of the network map based on the bibliographic data (e.g., keyword co-occurrences, citation, and bibliographic coupling).

It is clear from Figure 4 that the topic related to the energy efficiency and energy saving occurring in the research literature within the past five years often include optimization and design of the energy networks that are envisaged to optimize energy consumption and thermal comfort (cooling and heating) in the residential and (industrial) office buildings as well as in the (electric) transport.

Finally, we have singled out 91 papers from the sample of 12960 WoS publications using the additional keyword “COVID-19” and subjected the titles of these papers to the word analysis using the Word cloud diagram. The Word Cloud diagram shows the most frequent keywords, where a highlighted font yields higher frequency of using the term within the titles of publications retrieved from WoS and constituting parts of our sample.

In general terms, the results stemming from our bibliometric analysis confirm the findings achieved during the literature review conducted in the previous sections and yield similar implications and perspectives of the energy efficiency and energy saving for the post-COVID era.

## 5. CONCLUSIONS

Overall, the prospects for energy-saving technologies and energy efficiency in the post-COVID era are promising. By leveraging technological advancements, adopting effective policies, and promoting collaboration, societies can achieve sustainable energy practices. The lessons learned from the pandemic serve as a guide for developing resilient and flexible energy systems. Taking these prospects and recommendations on board might help to contribute to build a more sustainable future.

In the post-COVID era, as world’s societies and economies navigate the path to recovery and rebuilding, it is imperative that we prioritize energy-saving technologies and enhanced energy efficiency. The global health crisis offered us an unparalleled chance to introspect on our existing methods and implement significant alterations for a more sustainable future. This stems from the fact that energy-saving technologies hold a pivotal position in diminishing our carbon footprint and combatting climate change. By embracing these advancements, our society can significantly decrease its energy consumption while still meeting its needs. From smart grids and renewable energy sources to energy-efficient appliances and buildings, there are numerous ways in which our society can optimize its energy usage.

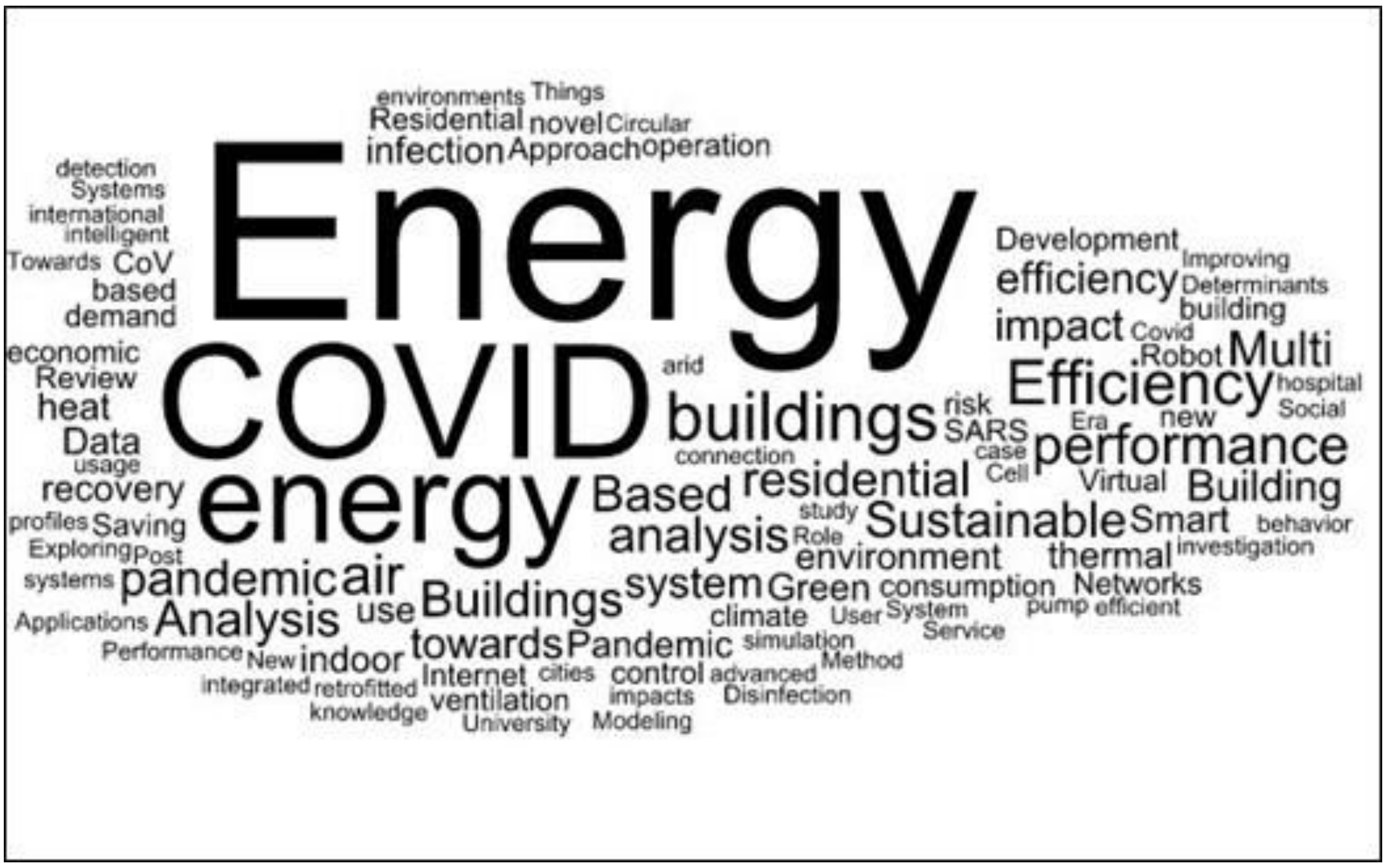

**Figure 5.** Word cloud diagram based on the titles of 91 publications indexed in WoS using the search terms “energy efficiency”, “energy saving”, and “COVID-19”.

Furthermore, directing investments towards cutting-edge clean and sustainable technologies leads to the creation of fresh employment prospects, fostering innovation and facilitating economic growth. To fully embrace a greener future, it is essential for governments, businesses, and individuals (consumers and prosumers) to collaborate. Within this context, governments should provide incentives for adopting sustainable practices while establishing stricter regulations on carbon emissions. Enterprises should give utmost importance to sustainability as an integral aspect of their corporate social responsibility endeavours. Likewise, individuals should receive education on the significance of energy efficiency and be motivated to make informed decisions that contribute to a more environmentally friendly world. Energy consumers should transition to becoming prosumers, actively generating, and trading their energy, utilizing innovative technological solutions like P2P networks, which will play a central role in shaping the smart grids of the 21st century.

Moreover, the COVID-19 pandemic brought about significant changes in the way energy-saving technologies and energy efficiency are perceived and regulated. In the wake of the COVID-19 crisis, governments worldwide struggled with its economic and environmental consequences. Energy-saving technologies are emerging because of this struggle as a crucial element in sculpting our future landscape. The pandemic removed many barriers for the legislation that is required for providing incentives for the adoption of these technologies (e.g. adapting novel building regulations, providing tax benefits for buying green technologies or introducing higher efficiency standards for electric vehicles). The recent developments demonstrated that energy efficiency and energy saving both help to tackle global warming and foster sustainable economic growth. As a result, governments worldwide would have to promote renewable energy sources as part of the post-COVID recovery strategy.

Regarding the above, the role of the international collaboration between governments and organizations seeking to achieve these priorities would become crucial. This cooperation would help to promote energy-saving technologies, increase the volume of technological transfers, as well as facilitate mutual learning and exchange of the best practices.

Our comprehensive literature review highlights the necessity for robust policy frameworks that incentivize energy efficiency, technological innovations for the seamless integration of energy-saving technologies, and public engagement to promote a shift towards sustainable energy usage. The environmental ramifications of energy consumption underscore the urgency of implementing energy-saving measures globally. Our goal is to galvanize policymakers, researchers, and industry experts to place energy efficiency and the adoption of energy-saving technologies at the forefront of strategies for a sustainable and resilient future, one that navigates the challenges and seizes the opportunities of the post-COVID era, marked by rapid technological progress.

**6. AUTHORS' NOTE**

The authors declare that there is no conflict of interest regarding the publication of this article. The authors confirmed that the paper was free of plagiarism.